\documentclass[aps,pra, twocolumn, superscriptaddress,amssymb,showpacs]{revtex4-1}
\usepackage{graphicx}
\usepackage{epstopdf}

\begin{document}
\title{Bistability and chaos at low-level of quanta}

\author{T.~V.~Gevorgyan}
\email[]{t\_gevorgyan@ysu.am}
\affiliation{Institute for Physical Researches, National Academy
of Sciences,\\Ashtarak-2, 0203, Ashtarak, Armenia}

\author{A.~R.~Shahinyan}
\email[]{anna\_shahinyan@ysu.am}
\affiliation{Yerevan State University, Alex Manoogian 1, 0025,
Yerevan, Armenia}

\author{Lock Yue Chew}
\email[]{lockyue@ntu.edu.sg}
\affiliation{Nanyang Technological University,
21 Nanyang Link, SPMS-PAP-04-04, Singapore 637371}

\author{G.~Yu.~Kryuchkyan}
\email[]{kryuchkyan@ysu.am}
\affiliation{Institute for Physical Researches,
National Academy of Sciences,\\Ashtarak-2, 0203, Ashtarak,
Armenia}\affiliation{Yerevan State University, Alex Manoogian 1, 0025,
Yerevan, Armenia}

\begin{abstract}
{We study nonlinear phenomena of bistability and chaos at a level of few quanta. For this purpose we consider a single-mode dissipative oscillator with strong Kerr nonlinearity with
respect to dissipation rate driven by a monochromatic force as well as by a
train of Gaussian pulses. The quantum effects and decoherence in oscillatory
mode are investigated on the framework of the purity of states and the Wigner
functions calculated from the master equation. We demonstrate the quantum
chaotic regime by means of a comparison between the contour plots of the Wigner
functions and the strange attractors on the classical Poincar\'e section.
Considering bistability at low-limit of quanta, we analyze what is the minimal
level of excitation numbers at which the bistable regime of the system is
displayed? We also discuss the formation of oscillatory chaotic regime by
varying oscillatory excitation numbers at ranges of few quanta. We demonstrate
quantum-interference phenomena that are assisted hysteresis-cycle behavior and
quantum chaos for the oscillator driven by the train of Gaussian pulses as well
as we establish the border of classical-quantum correspondence for chaotic
regimes in the case of strong nonlinearities.}
\end{abstract}

\maketitle

\section{Introduction}\label{intro}

Nonlinear dissipative oscillator (NDO) operated in the quantum
regime is becoming significant in both fundamental and applied sciences,
particularly, for implementation of basic quantum optical systems, in
engineering of nonclassical states and quantum logic. The important
implementations have recently been realized in the context of superconducting
devices based on the nonlinearity of the Josephson junction (JJ) exhibiting a
wide variety of quantum phenomena (see, the Reviews \cite{Nori}, \cite{Nori1}
and \cite{Mak}-\cite{Hosk}). In some of these devices, dynamics are analogous to
those of a quantum particle in an oscillatory anharmonic potential
\cite{claud}. The single nonlinear oscillator and systems of nonlinear
oscillators also consist of basic theoretical models for various
nano-electro-mechanical and nano-opto-mechanical devices. In the last decade
there was exciting technological advances in the fabrication and control of
such devices \cite{Craig}-\cite{ekinci} that are attracting interest in a broad
variety of research areas and for many possible applications due to their
remarkable combination of properties: small mass, high operating frequency,
large quality factor, and easily accessible nonlinearity. Nanomechanical
oscillators are being developed for a host of nanotechnological applications.
They are ideal candidates for probing quantum limits of mechanical motion in an
experimental setting. Moreover, they are the basis of various precision
measurements \cite{Jens}-\cite{Rug17}, as well as for basic research in the
mesoscopic physics of phonons \cite{Keith18}, and the general study of the
behavior of mechanical degrees of freedom at the interface between the quantum
and the classical worlds \cite{Miles}.

The efficiency of quantum oscillatory effects requires a high nonlinearity with
respect to dissipation. However, for weak damping even small nonlinearity can
become important. Development of driven  NDO in quantum regime requires
cooling
these systems to their ground state and significant advances have been made in
cooling the systems to far below the temperature of the environment \cite{Rae}-\cite{Wang26}.

It is well known that classically, the nonlinearity-induced dependence of the
oscillatory frequency or the amplitude usually leads to bistability of the driven NDO \cite{drummond}. The corresponding systems describe amplifiers which are
ubiquitous in experimental physics. The Josephson junction amplifier has been
discussed in \cite{JBA} as a bistable amplifier which has particular interest.
The dynamical bifurcation of a rf-biased Josephson junction was proposed to be
used as a basis for the amplification of quantum signals in \cite{vij1, vij2}.
Though the bistability has been usually treated as a classical signature of the NDO
the quantum dynamics in the bistable region has been a new subject in the past
years \cite{ep1}-\cite{ep4}.

Quantum dynamics of an oscillator is naturally described by Fock states, that
have definite numbers of energy quanta. However, these states are hard to
create in experiments because excitations of oscillatory systems usually lead
to the production of coherent states but not quantum Fock states. Nevertheless,
quantum oscillatory states can be prepared and can be manipulated by coupling
oscillators to atomic systems driven by classical pulses. The systematic
procedure has been proposed in Ref. \cite{law} and has been demonstrated for
deterministic preparation of mechanical oscillatory Fock states with trapped
ions \cite{mek}, in cavity QEDs with Rydberg atoms \cite{varc} and in
solid-state circuit QED for deterministic preparation of photon number states
in a resonator by interposing a highly nonlinear Josephson phase qubit between
a superconducting resonator \cite{hof}.

For the NDO in quantum regime the nonlinearity makes frequencies of transitions
between adjacent oscillatory energy levels different. Thus, strong nonlinearity
enables spectroscopic identification and selective excitation of transitions
between Fock states. Thus, it is possible in this regime to prepare the NDO at
low-level of quanta. In this approach, it has been shown that the production of
Fock states, Fock states superpositions or qubits can also be realized in
over-transient regime of an anharmonic dissipative oscillator without any
interactions with atomic and spin-1/2 systems and with complete consideration
of decoherence effects \cite{Gev39}. For this goal the strong Kerr
nonlinearity as well as the excitation of resolved lower oscillatory energy
levels with a specific train of Gaussian pulses have been considered.

In this paper we consider the NDO in the regime of low-level of excitation for the study of
the problems of quantum bistability and chaos. Therefore, the goal of the paper
is twofold. In one part, we consider the bistability on a few oscillatory
excitation number considering the NDO driven by monochromatic force. The
bistability on a few excitation number is attractive for ultra-low power
operation, but it has practical problems related with quantum
fluctuation-induced spontaneous switching. In this part, we also demonstrate
the production of quantum interference between bistable branches for the NDO
driven by a train of Gaussian pulses.

The other part of the paper is devoted to investigation of quantum chaos in
low-level excitation regime of the pulsed NDO. Much research on the subject of
classical and quantum chaos has been done on the base of the kicked rotor,
which exhibits regions of regular and chaotic motion. Its experimental
realization, and observation of the model’s dissipation and decoherence
effects are carried out on a gas of ultracold atoms in a magneto-optical trap
subjected to a pulsed standing wave \cite{Ammann41}, \cite{Klappauf42}. In Ref.
\cite{Milburn43} it was proposed to realize the parametrically kicked nonlinear
oscillator model in a cavity involving Kerr nonlinearity. It was also shown
that a more promising realization of this system, including the quantum
regime, is achieved in the dynamics of cooled and trapped ions, interacting
with a periodic sequence of both standing wave pulses and Gaussian laser pulses
\cite{Breslin44}. In addition to these  and the other important input (see, also \cite{Leon}, \cite{Macc})  in this paper we consider quantum dissipative chaos of the NDO driven by a train of Gaussian
pulses in a strong quantum regime and in complete consideration of dissipation
and decoherence. In this way, quantum dissipative chaos at the limit of low-level
of excitation numbers will be considered.

The paper is arranged as follows. In Sec. II, we shortly describe a pulsed NDO.
In Sec. III, we study bistability at a level of few quanta for the NDO driven by
monochromatic force. In Sec.IV, we consider the production of quantum
interference for bistable regime of the NDO driven by the train of Gaussian pulses.
In Sec. V, we study quantum dissipative chaos at limit of low-level of
excitation numbers for the NDO driven by the train of Gaussian pulses. We
summarize our results in Sec. VI.

\section{ The model: short description}
The Hamiltonian of an anharmonic-driven oscillator in the rotating-wave approximation reads as:
\begin{equation}
H=\hbar \Delta a^{\dagger}a + \hbar \chi (a^{\dagger}a)^{2} +
\hbar f(t)(\Omega a^{\dagger} + \Omega^{*}a).\label{hamiltonian}
\end{equation}
Note that time dependent coupling constant $\Omega f(t)$ that is proportional to
the amplitude of the driving field consists of  Gaussian pulses with
duration $T$ which are separated by time intervals $\tau$ as follow
\begin{equation}
f(t)=\sum{e^{-(t - t_{0} - n\tau)^{2}/T^{2}}}. \label{driving}
\end{equation}
Here, $a^{\dagger}$, $a$ are the oscillatory creation and annihilation operators,
$\chi$ is the nonlinearity strength, and $\Delta=\omega_{0} -\omega$ is the
detuning between the mean frequency of the driving field and the frequency of
the oscillator. For $f(t)=1$ this Hamiltonian describes nonlinear oscillator driven by a monochromatic force.

The evolution of the system of interest is governed by
the following master equation for the reduced density matrix
in the interaction picture:

\begin{equation}
\frac{d\rho}{dt} =-\frac{i}{\hbar}[H, \rho] +
\sum_{i=1,2}\left( L_{i}\rho
L_{i}^{\dagger}-\frac{1}{2}L_{i}^{\dagger}L_{i}\rho-\frac{1}{2}\rho L_{i}^{\dagger}
L_{i}\right)\label{master},
\end{equation}
where $L_{1}=\sqrt{(N+1)\gamma}a$ and $L_{2}=\sqrt{N\gamma}a^{\dagger}$ are the
Lindblad operators, $\gamma$ is a dissipation rate, and $N$ denotes the mean
number of quanta of a heath bath. To study the pure quantum effects we focus below on the cases of very low reservoir
temperatures which, however, ought to be still larger than the characteristic
temperature $T \gg T_{cr}=\hbar\gamma/k_B$.

The Hamiltonian (\ref{hamiltonian}) describes wide range of physical systems,
including nano-mechanical oscillator, Josephson junction device, optical
fibers, quantum dots, quantum scissors. Part of them have been noted in section \ref{intro}. Note, that quantum effects in a NDO with a time-modulated driving force including also pulsed regime have been studied
in a series of papers \cite{qsch}-\cite{mpop}.

Below, we use numerical simulation of this equation based on quantum state diffusion method (QSD) (see, Ref. \cite{qsd} and, for example, applications in Refs.\cite{qsch}-\cite{mpop} and \cite{qsd1}-\cite{qsd4}).
For clarity, in our numerical
calculation we choose the mean number of reservoir photons $N=0$. Note, that for
$N\ll1$ the above mentioned restriction is valid for the majority of problems
of quantum optics and, particularly, for the schemes involving the nano-mechanical oscillator and Josephson junction.
In experiments, the nonlinear oscillator based on the current-biased JJ is
cooled down to $T=20mK$, which corresponds to $N=0.0013$,
whereas, $T_{cr}=10^{-5}$ K, for $\gamma=1 MHz$.

In semiclassical approach the corresponding equation of motion
for the dimensionless amplitude of oscillatory mode has the following form

\begin{equation}
\frac{d\alpha}{dt}= -i[\Delta + \chi + 2|\alpha|^2\chi]\alpha + if(t)\Omega -\gamma\alpha.\label{semclass}
\end{equation}
This equation modifies the standard Duffing equation on the case of the NDO
with time-dependent coefficient.

\section{Bistability at level of a few quanta}
At first, we describe the NDO driven by monochromatic force [the case, $f(t)$=1
in the Hamiltonian described by (1)] in bistable regime. In semiclassical
approach based on Eq. 4 the bistable dynamics is realized if the following
inequalities are satisfied \cite{drummond}:
\begin{eqnarray} \chi(\Delta + \chi) < 0, ~~~~~~~~~~~~~~~~~\nonumber \\
|(\Delta+\chi)/0.5\gamma| > \sqrt3, ~~~~~~~~~~~~~~~~\nonumber \\
 \left[ 1 + \frac{27 \chi \Omega^{2}}{\left( \Delta + \chi\right) ^{3}}+ \left( \frac{1.5 \gamma}{\Delta + \chi}\right) ^{2}\right] ^2 < \left[1 - 3\left( \frac{\gamma/2}{\chi + \Delta}\right) ^2\right] ^3 .\label{bistable_ineq}
\end{eqnarray}

\begin{figure}
\includegraphics[width=7.0cm]{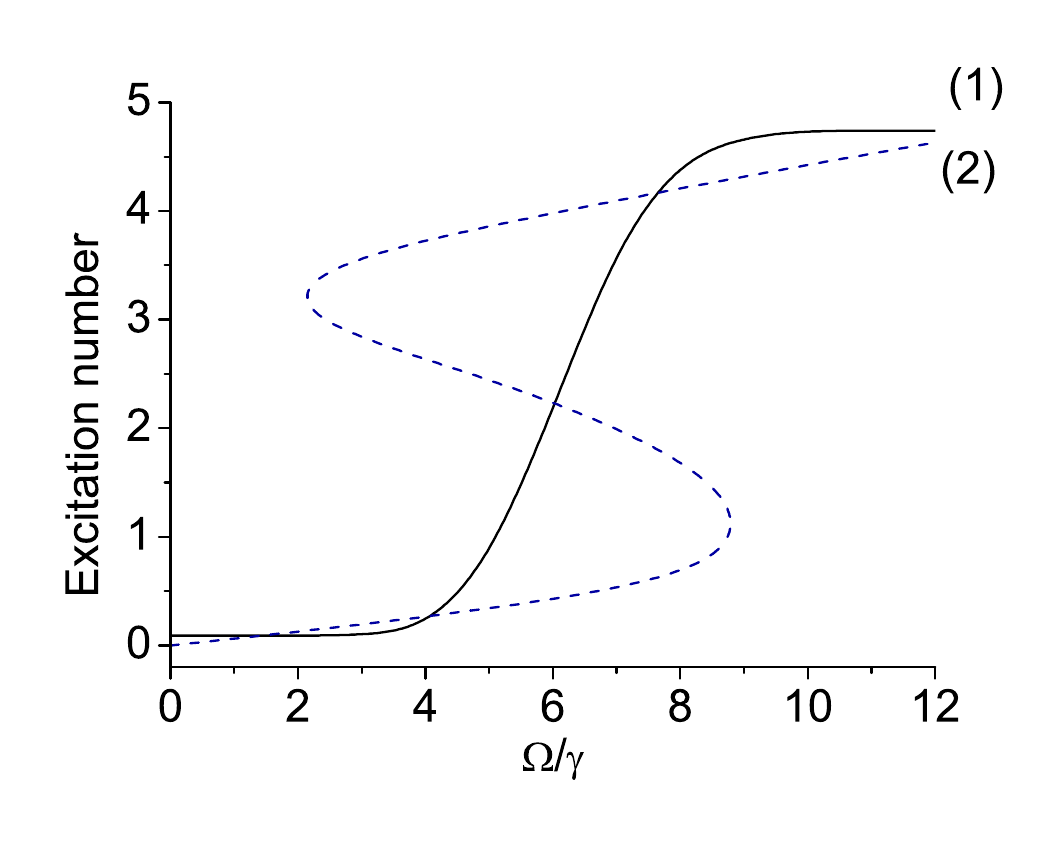}
\caption{(Color online) The mean excitation numbers of anharmonic oscillator: (1) quantum solution, (2) semiclassical solution. The parameters are as follows:
$\Delta/\gamma = -15$, $\chi/\gamma=2$.}
\label{hyster}
\end{figure}

In this range of the parameters the typical hysteresis curves depending
on detuning $\Delta$ and force amplitude $\Omega$ display the system \cite{drummond}. This result for the stationary excitation number $n=|\alpha|^2$ can be obtained by solving the following equation:
\begin{equation}
|\alpha|^2=\frac{\Omega^2}{\left(\Delta+\chi+2\chi|\alpha|^2\right)^2+\left(\gamma/2\right)^2}
\label{semi}
\end{equation}
and is depicted in Fig. \ref{hyster}.

It is well known that whereas the semiclassical result exhibits
hysteresis-cycle behavior,
the corresponding quantum mechanical result, which
accounts for the influence of quantum noise, shows a
gradual evolution. It is seen from the exact quantum solution for the
mean excitation number in the following form:
\begin{equation}
\langle a^{\dagger}a\rangle = \frac{\Omega^2}{(\Delta + \chi)^2 + (\gamma/2)^2} \frac{F(c + 1, c^{*} + 1, z)}{F(c, c^{*}, z)},
\label{quant}
\end{equation}
where $F = F_{2}$ is the generalized hypergeometric function, $c$ and $z$  are coefficients that depend on the parameters: $c = (\Delta + \chi)/\chi - i\gamma/(2\chi)$
and $z = 2(\Omega/\chi)^2$.

It is also seen that the characteristic
threshold behavior, determined by a drastic increase
of the intensity in the transition region, disappears
as the relative nonlinearity $\chi/\gamma$ increases.
In Fig. \ref{hyster} we plot both quantum and semiclassical solutions
corresponding to Eqs. \ref{semi},\ref{quant}.

More detailed information on the quantum statistical
properties of the oscillatory mode in the bistable range can be obtained
from the analysis of the excitation number probability
distribution function $p(n)$. In this way, the locations of extrema
of the $p(n)$-function, i.e. the locations of the
most and least probable values of $n$, may be identified
with the semiclassical stable and unstable steady
states in the limit of small quantum noise level \cite{a33}. With the increase of $\chi/\gamma$ the curve of locations of these
extrema as depending on $\Omega$ becomes shifted from
the corresponding semiclassical curve for the mean excitation number.

Considering bistability at low-limit of quanta, for strong quantum regime, it
is naturally to put the question what is the minimal level of excitation numbers
at which bistable regime is displayed? Below, we discuss this problem on the
base of quantum trajectories and the Wigner function to discuss quantum
dynamics and on the Poincar\'e section to discuss semiclassical
dynamics.

Analyzing the monochromatically driven NDO for strong quantum regime and in
over transient time intervals, $t\gg\gamma^{-1}$, we use numerical method but
not the analytical results obtained in terms of the exact solution of the
Fokker-Planck equation \cite{a33}- \cite{i34}. The reason is that the
steady-state solution of the Fokker-Planck equation has been found using the
standard approximation method of potential equations. The validity of this
solution has not been checked up to now in the strong quantum regime that
requires a high nonlinearity with respect to dissipation.

We use the Wigner function
\begin{equation}
W(r,\theta)=\sum_{n,m}\rho_{nm}(t)W_{mn}(r,\theta)
\end{equation}
in terms of the matrix elements
$\rho_{nm}=\langle n|\rho|m\rangle$ of the density matrix operator in the Fock
state representation. Here $(r,\theta)$ are the polar coordinates in the
complex phase space plane, $x=r\cos\theta$, $y=r\sin\theta$, while the
coefficients $W_{mn}(r,\theta)$ are the Fourier transform of matrix elements of
the Wigner characteristic function.

\begin{figure}
\includegraphics[width=8.6cm]{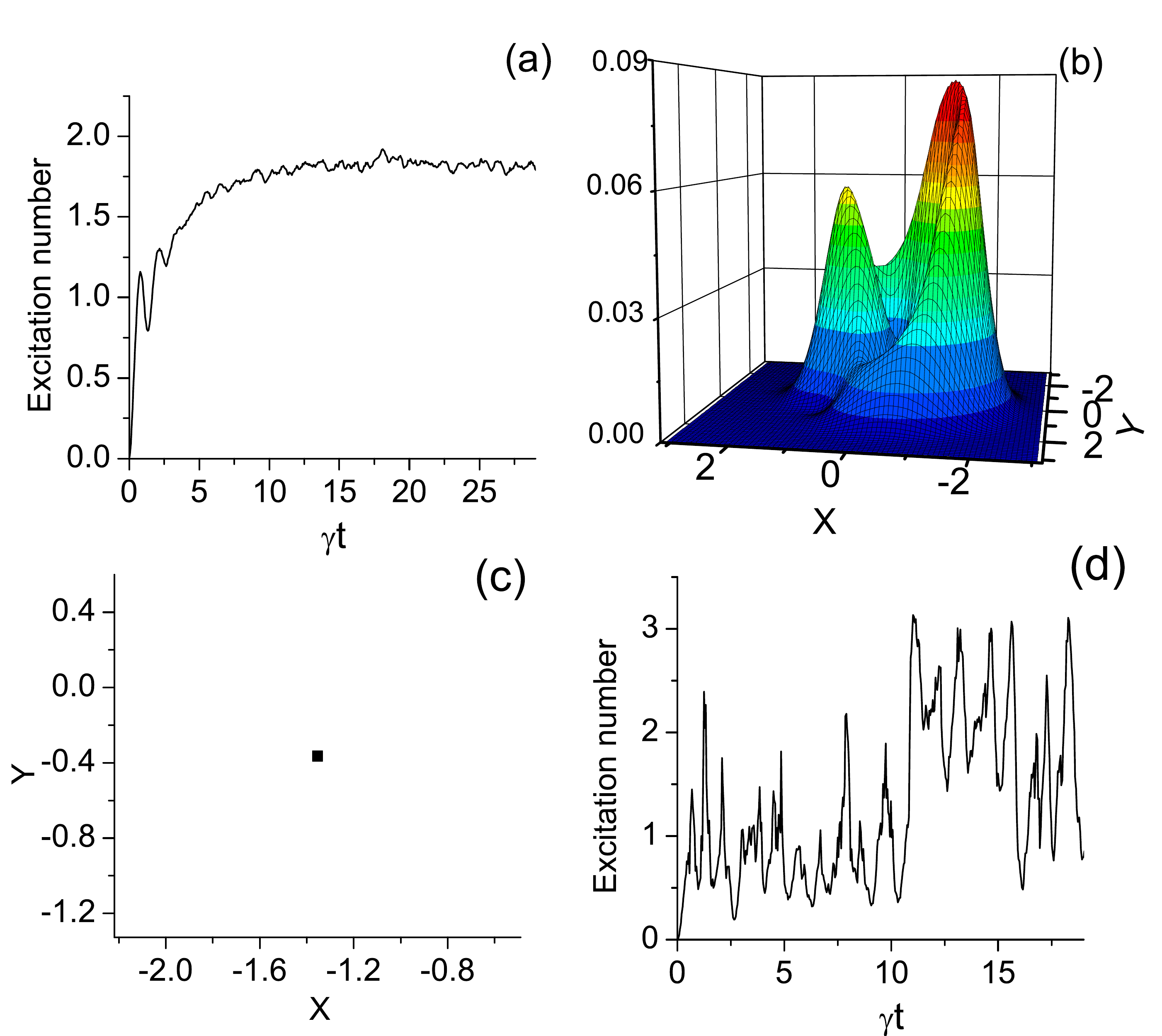}
\caption{(Color online) The mean excitation numbers of oscillatory mode (a),
 the Wigner function (b), semiclassical Poincar\'e section (c) , the time evolution of the excitation numbers along a single trajectory (d).
The parameters are as follows:
$\Delta/\gamma = -8$, $\chi/\gamma=2$, and $\Omega/\gamma = 2.7$}
\label{monochrom_bistable}
\end{figure}

The properties of bistable dynamics at the level of a few excitation number are
demonstrated on Fig.~\ref{monochrom_bistable}. 
Fig.~\ref{monochrom_bistable}(a) shows that the mean excitation number is small
which means that the system is operated in a deep quantum regime. Analyzing one
single quantum stochastic trajectory for excitation numbers on
Fig.~\ref{monochrom_bistable}(d) we set the system initially to the vacuum
oscillatory state and consider time-dependence over a long time, compared to
the characteristic dissipative time. As expected, the analysis of the
time-dependent stochastic trajectories for an expectation number shows that the
system spends most of its time close to one of the semiclassical bistable
solutions with quantum interstate transitions, occurring at random intervals.

In order to demonstrate the bistability in phase-space we tune the nonlinear
oscillator parameters to satisfy inequalities Eq.~(\ref{bistable_ineq}). As
calculations show the Wigner function displays two peaks
(Fig.~\ref{monochrom_bistable}(b)) that displays bistability, whereas
Poincar\'e section obtained from the semi-classical calculations of
Eq.~(\ref{semclass}) shows a single point in phase space (Fig.~ \ref{monochrom_bistable}(c)) that corresponds to the regular dynamics.

\begin{figure}
\includegraphics[width=8.6cm]{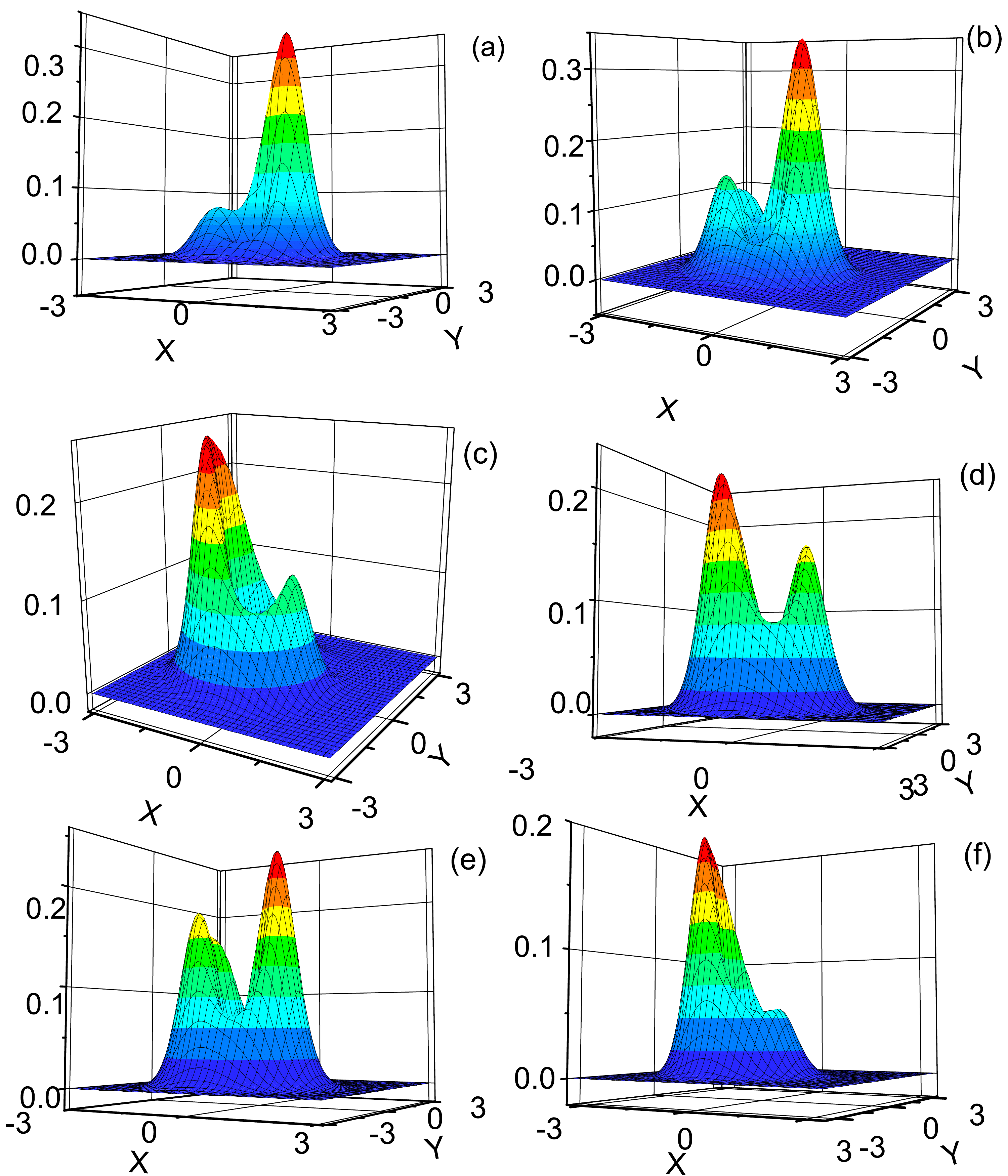}
\caption{(Color online) The Wigner functions for oscillatory mode in dependence on the amplitude. The parameters are as follows:
$\Delta/\gamma = -8$, $\chi/\gamma=2$, and (a) $\Omega/\gamma = 2.1$, (b) $\Omega/\gamma = 2.3$, (c) $\Omega/\gamma = 2.5$, (d) $\Omega/\gamma = 2.7$, (e) $\Omega/\gamma = 2.9$, (f) $\Omega/\gamma = 3.1$.}
\label{dynbistable}
\end{figure}

Further investigation of this model allows to establish
other properties of dissipative bistable dynamics. For this goal on the Fig.~\ref{dynbistable} we show bistability in phase-space depending on the amplitude of the external force. As we see for the given parameters of detuning and nonlinearity there is an intermediate range of amplitudes where bistability takes place.

\begin{figure}
\includegraphics[width=8.6cm]{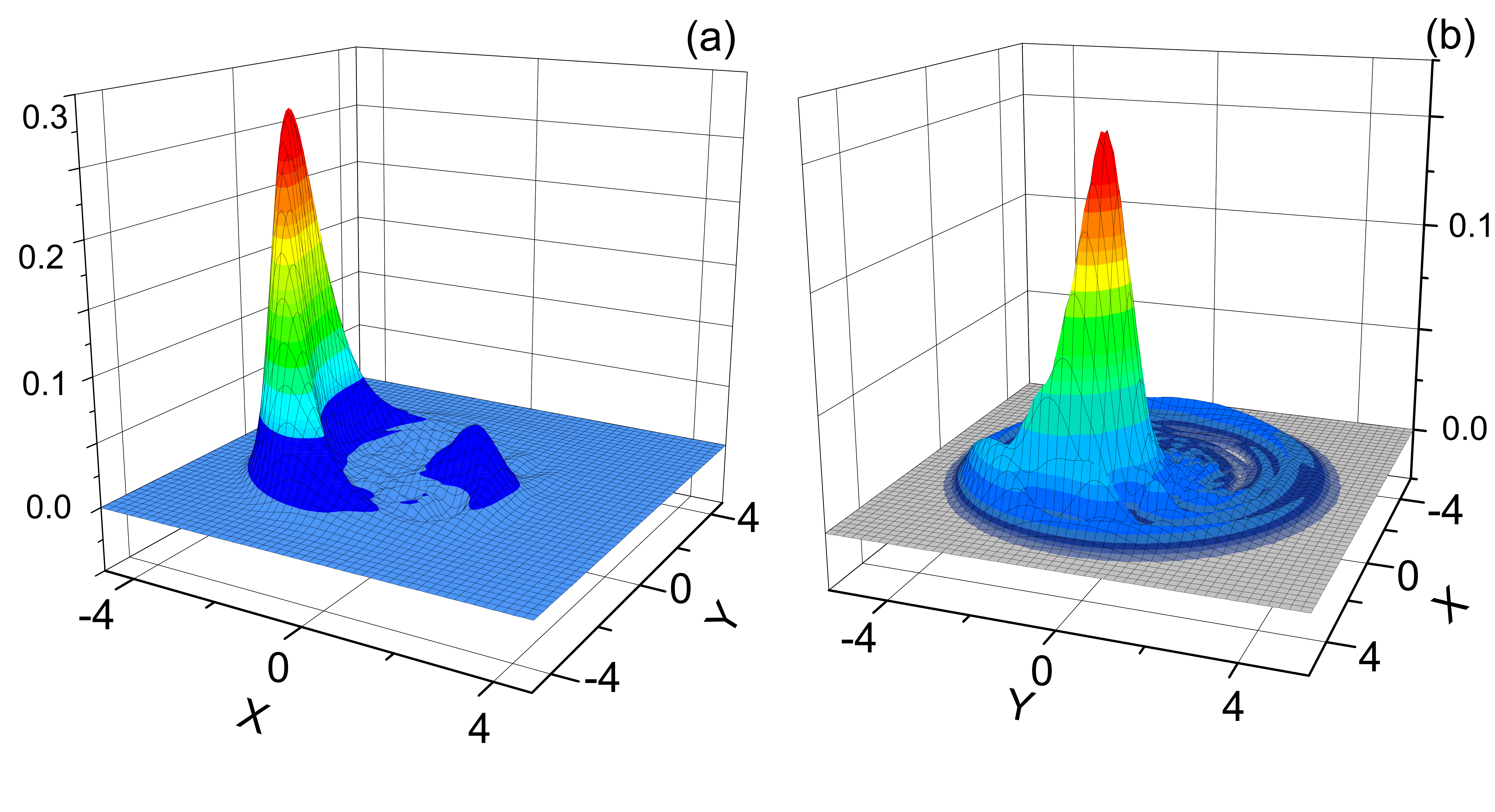}
\caption{(Color online) (a) The Wigner function for $\lambda = 2$, (b) the Wigner function
for $\lambda = 3$. The parameters are as follows: $\Delta/\gamma = -8$, $\chi/\gamma=2$, and $\Omega/\gamma = 2.7$}
\label{scaledW}
\end{figure}

It is also interesting to consider behavior of the NDO
 by using the scaling
properties of Eq.~(\ref{semclass}). Indeed, it is easy to
verify that this equation is invariant with respect to the following
scaling transformation of the complex amplitude: $\alpha^{'}\rightarrow \lambda\alpha, \chi^{'}\rightarrow\chi/\lambda^2,
\Omega^{'}\rightarrow\lambda\Omega, \Delta^{'}\rightarrow\Delta + \chi(1-1/\lambda^{2})$.
Thus, for $\lambda > 1$ oscillatory excitation numbers are increased during
scaling transformation. It is interesting to analyze such scaling from the
point of view of quantum-statistical theory and its relevance for decoherence
and dissipation. Thus, using the scaling properties we consider the system for
various excitation numbers. The Wigner functions for the scaled $\lambda = 2$
and $\lambda = 3$ are presented in Fig. \ref{scaledW}. As we see, increasing
$\lambda$ leads to the suppression of one of the peaks in the Wigner function and on the
whole the bistability of the system is vanished. Thus, we show that such
parameter scaling does not occur for strong quantum regime of NDO.

There is a boundary to detect bistable states at limit of small excitation
numbers in phase-space in accordance with Planck uncertainty principle. Taking
into account that $\Delta X \Delta P \geq 1/2$, where $X$ and $P$ are
dimensionless position and momentum operators and $n=|\alpha|^2$, $\alpha = X +
iP$, it seems that for small enough  level of excitation numbers we cannot
distinguish two branches of bistability.  In this case sizes of contour plots
of Wigner functions are sufficiently squeezed such that  two bistable branches are too
close to each other to be distinguishable.

\begin{figure}
\includegraphics[width=8.6cm]{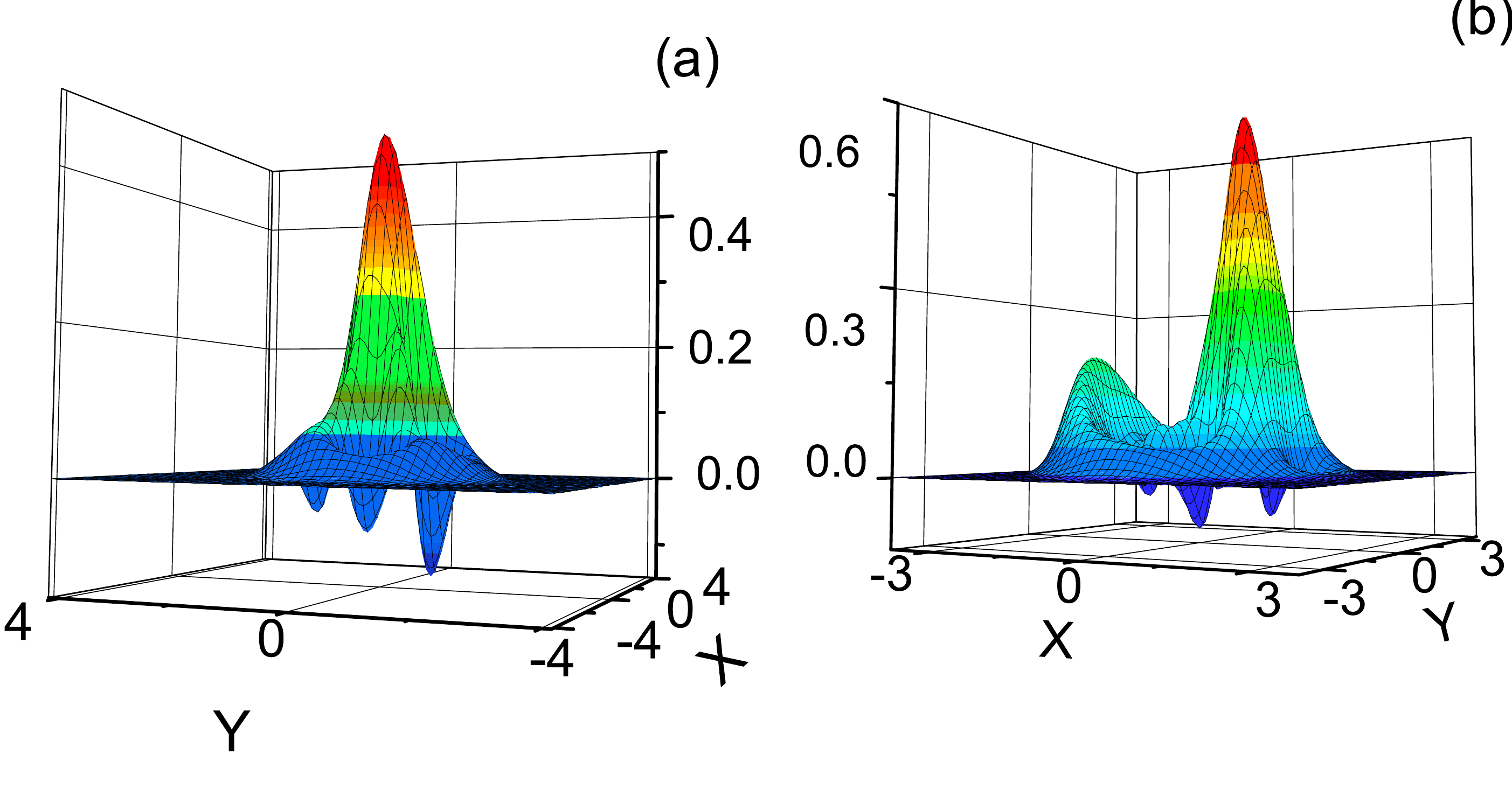}
\caption{(Color online) The Wigner functions showing quantum interference fringes. The parameters are as follows:
$\Delta/\gamma = -8$, $\chi/\gamma=2$, $\Omega/\gamma=2.7$, (a) $T=0.5\gamma^{-1}$ $\tau=2\gamma^{-1}$, (b) $T=0.1\gamma^{-1}$, and $\tau=2\gamma^{-1}$}.
 \label{wignermodulated}
\end{figure}
\begin{figure}
\includegraphics[width=8.6cm]{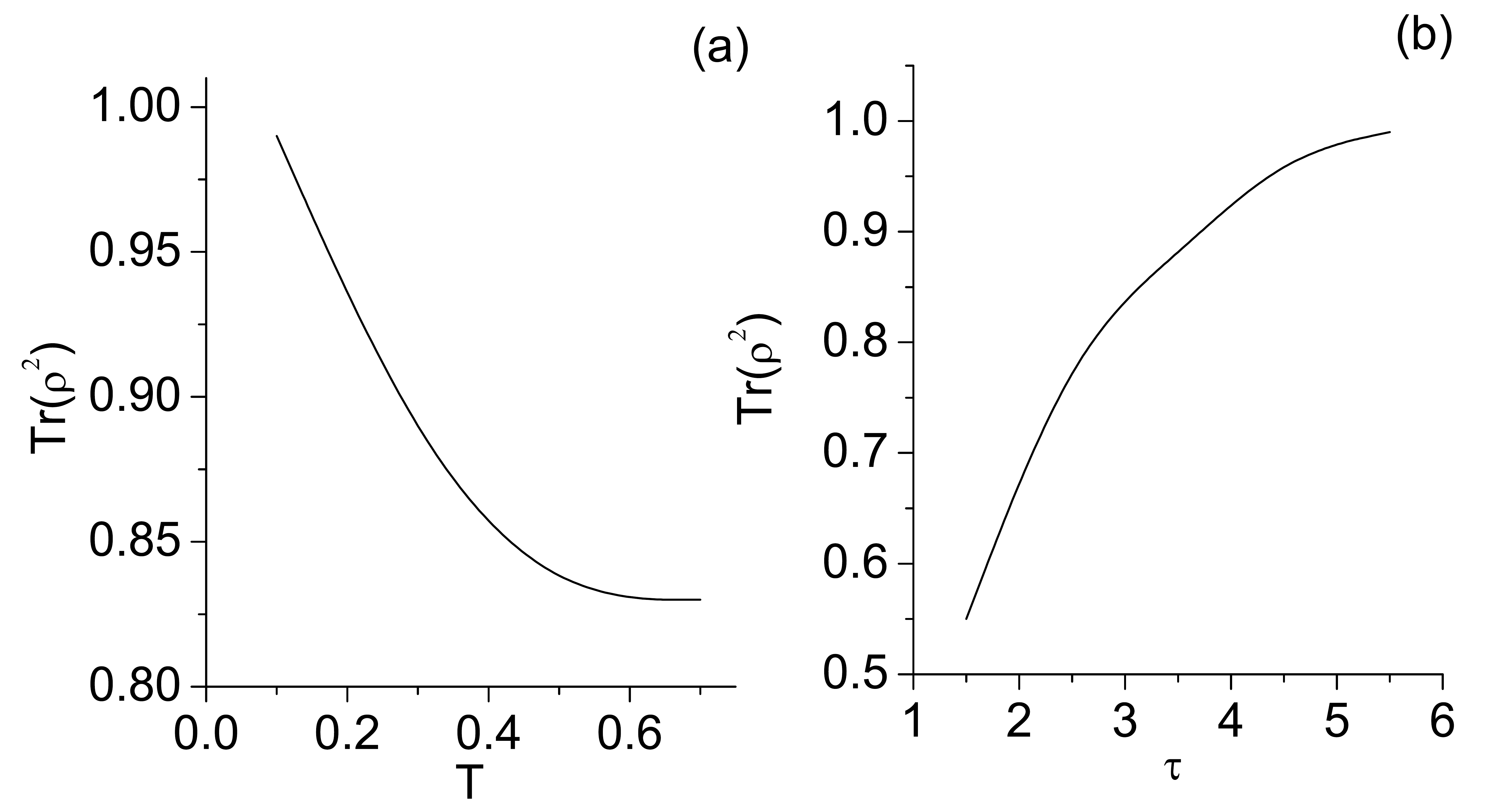}
\caption{The purity depending from $T$ for fixed $\tau=2.5\gamma^{-1}$ (a), the purity depending from $\tau$ for fixed $T=0.5\gamma^{-1}$ (b). The other parameters are: $\Delta/\gamma = -8$, $\chi/\gamma=2$, and $\Omega/\gamma=2.7$.}
\label{purity_dep}
\end{figure}

\section{Quantum interference assisted by bistability}
In this section we demonstrate that it is possible to create quantum
superposition in bistable dynamics of the NDO under pulsed excitation. Really, the
application of time-dependent force can lead to transition between two
branches of the system dynamics in the bistable regime and open an opportunity
to generate interference between them. However, quantum interference takes
place for very short time intervals and disappears due to dissipation and
decoherence. In order to recover the quantum interference for over transient
regimes we suggest to apply the specific train of Gaussian pulses according to
the model considered in  section II.
The results depicted on  Fig. \ref{wignermodulated} shows that with applied
Gaussian pulse the Wigner function has negative ranges. The Wigner functions
describe two humps corresponding to bistable branches and interference pattern
between them in phase-space.

In order to be sure that effects of dissipation and decoherence in oscillatory mode for transient time are suppressed if the train of pulses is applied we
calculate the purity of the state, i.e. $Tr(\rho^2)$. For a pure state
$Tr(\rho^2) = 1$. The results for $Tr(\rho^2)$ are depicted in Fig. \ref{purity_dep} for over-transient regime, i.e. for time intervals, $t\gg\gamma^{-1}$, in dependence on the pulse duration $T$ and time-intervals between pulses $\tau$. The
Fig. \ref{purity_dep}(a) corresponds to the case of fixed $\tau$ , i.e. $\tau=2.5$. Considering the dependence from the duration of pulses for this case we conclude that the purity is maximal for very short pulses and lost with increasing of $T$. The opposite behavior for the fixed interval between pulses is realized in Fig. \ref{purity_dep}(b).

\section{Quantum dissipative chaos at low-level of excitation numbers}
In this section we demonstrate that dissipative chaos is realized in strong
quantum regime NDO on low-level of oscillatory excitation numbers. Chaotic
regime appears in the NDO driven by train of Gaussian pulses, and it depends on
the duration $T$ of the pulses and the time intervals $\tau$ between them.

Many criteria have been suggested to define chaos in quantum systems, varying
in their emphasis and domain of application. Nevertheless, as yet, there is no
universally accepted definition of quantum chaos. Our analysis is given in the
framework of semiclassical and quantum distributions by using a correspondence
between contour plots of the Wigner function and the Poincar\'e section. Such
analysis has been proposed and realized \cite{AMK1} in mesoscopic regimes of
NDO with time-dependent coefficients.

It is well known that the Poincar\'e section has the form of strange attractor
in phase space for dissipative chaotic systems while it has the form of close
contours with separatrices for Hamiltonian systems. Thus, in this paper we
demonstrate the quantum chaotic regime by means of a comparison between the
contour plots of the Wigner functions and the strange attractors on the
classical Poincar\'e section. In this way, we calculate the Wigner function in
phase space by averaging an ensemble of quantum trajectories for definite time
intervals. On the other hand, the Poincar\'e section is calculated through the
semiclassical distribution based on Eq. \ref{semclass} but for a large number
of time intervals: it is constructed by fixing points in phase space at a
sequence of periodic intervals. Note, that such analysis seems to be rather
qualitative than quantitative for strong quantum regime and the ranges of
low-level oscillatory excitation numbers, where the validity of semiclassical
equation is questionable. Indeed, it is shown below that semiclassical and
quantum treatment of quantum dissipative chaos are cardinally different in the
deep quantum regime.

\begin{figure}
\includegraphics[width=8.6cm]{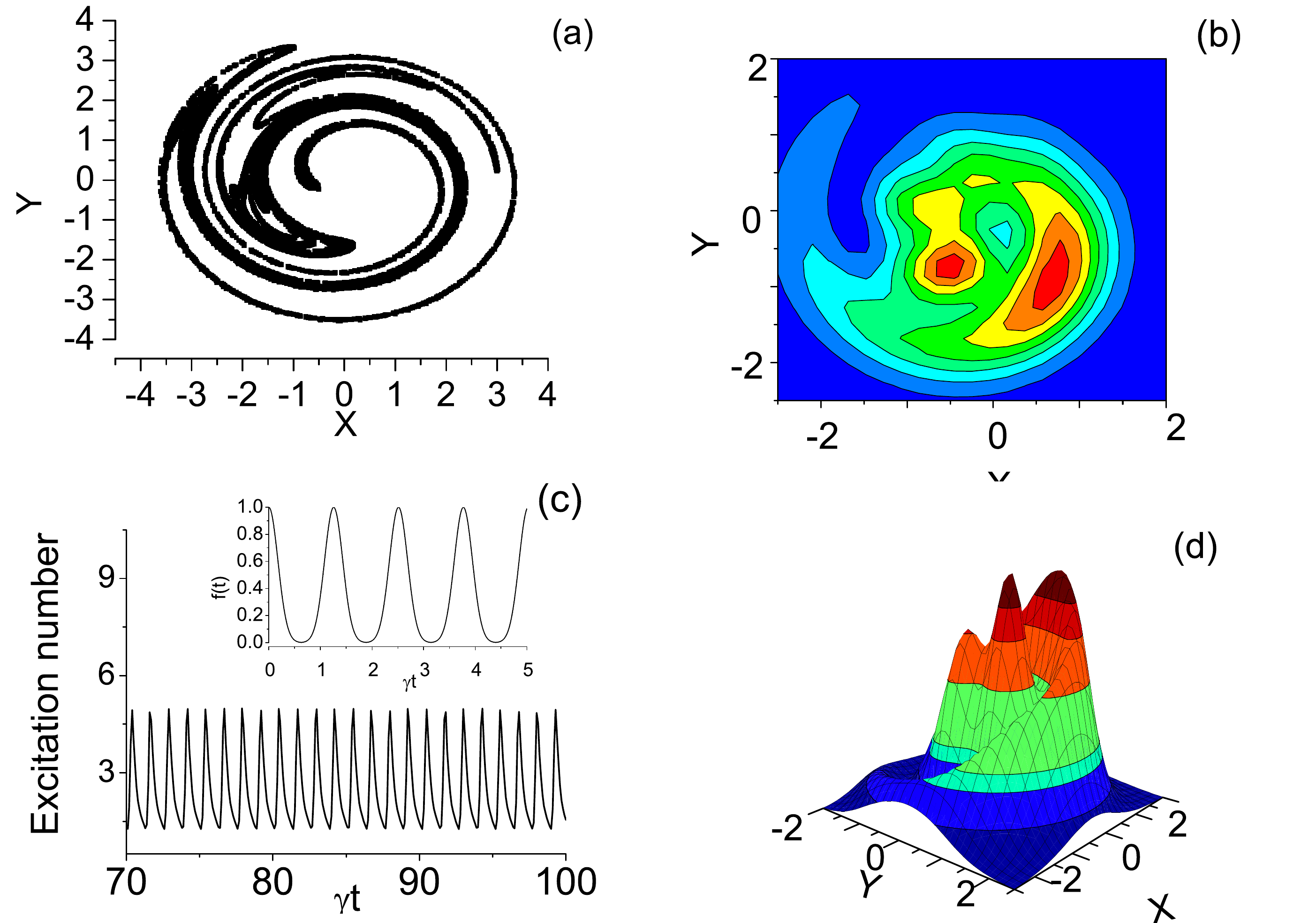}
\caption{(Color online)(a)The Poincar\'e section, (b)the contour plot of the Wigner function, (c)the excitation number time series with snapshots of Gaussian pulses as inset, and (d) the Wigner function. The parameters are as follows: $\chi/\gamma=0.7$, $\Omega/\gamma=20.4$, $\Delta/\gamma=-15$, $T=0.25\gamma^{-1}$, $\tau=2\pi/5\gamma$. Note that these distributions occur at time $\gamma t=100$ with the mean excitation number of $1.54$. The range of excitation numbers are $1.26$ to $4.98$.}
\label{pulse_duration1}
\end{figure}

\begin{figure}
\includegraphics[width=8.6cm]{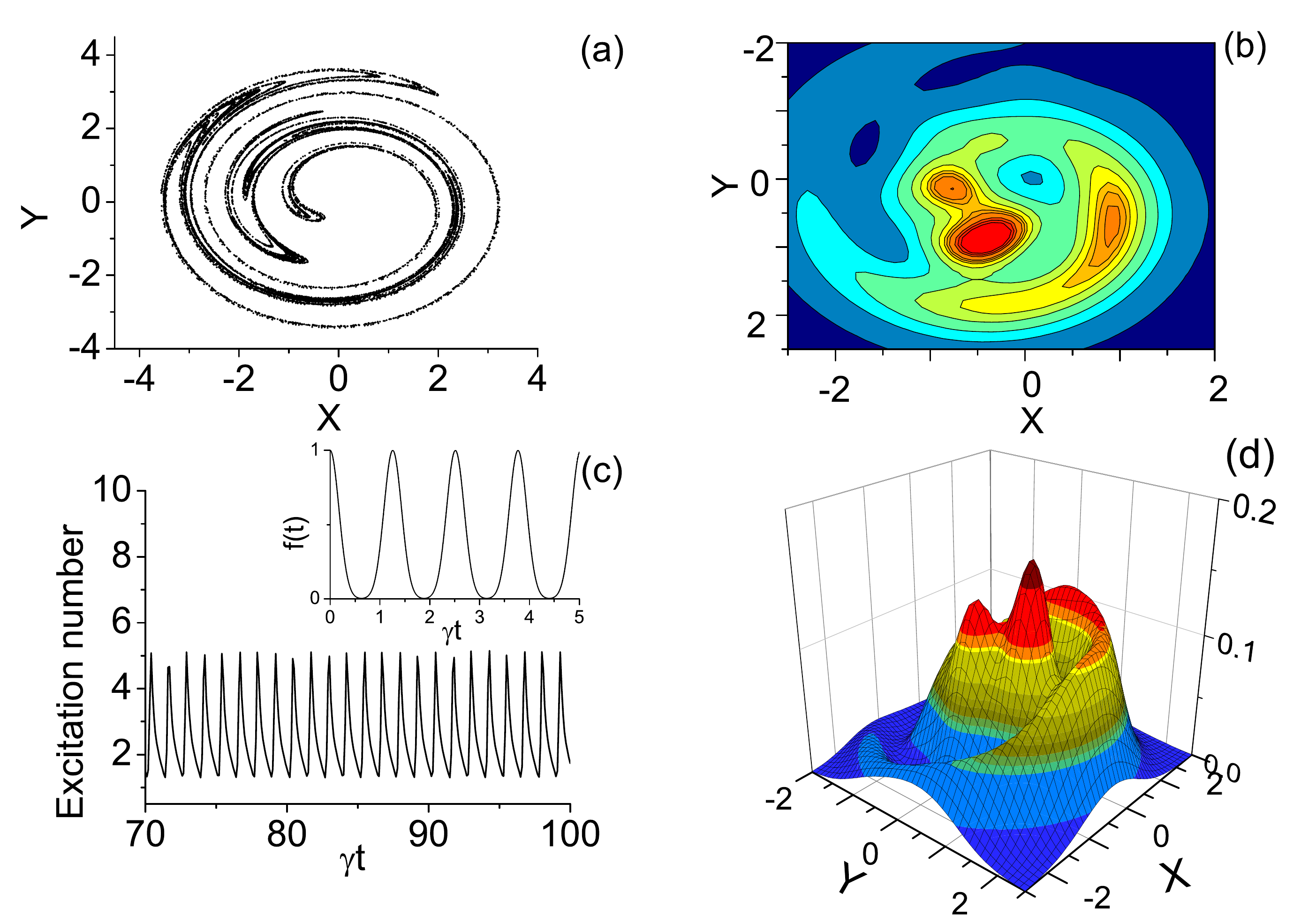}
\caption{(Color online)(a) The Poincar\'e section, (b) the contour plot of Wigner function, (c) the excitation number time series with snapshots of Gaussian pulses as inset, and (d)the Wigner function. The parameters are: $\chi/\gamma=0.7$, $\Omega/\gamma=20.4$, $\Delta/\gamma=-15$, $T=0.205\gamma^{-1}$, $\tau=2\pi/5\gamma$. Note that these distributions occur at time $\gamma t=100$ with the mean excitation number of $1.74$. The range of excitation numbers are $1.29$ to $5.11$.}
\label{pulse_duration2}
\end{figure}

\begin{figure}
\includegraphics[width=8.6cm]{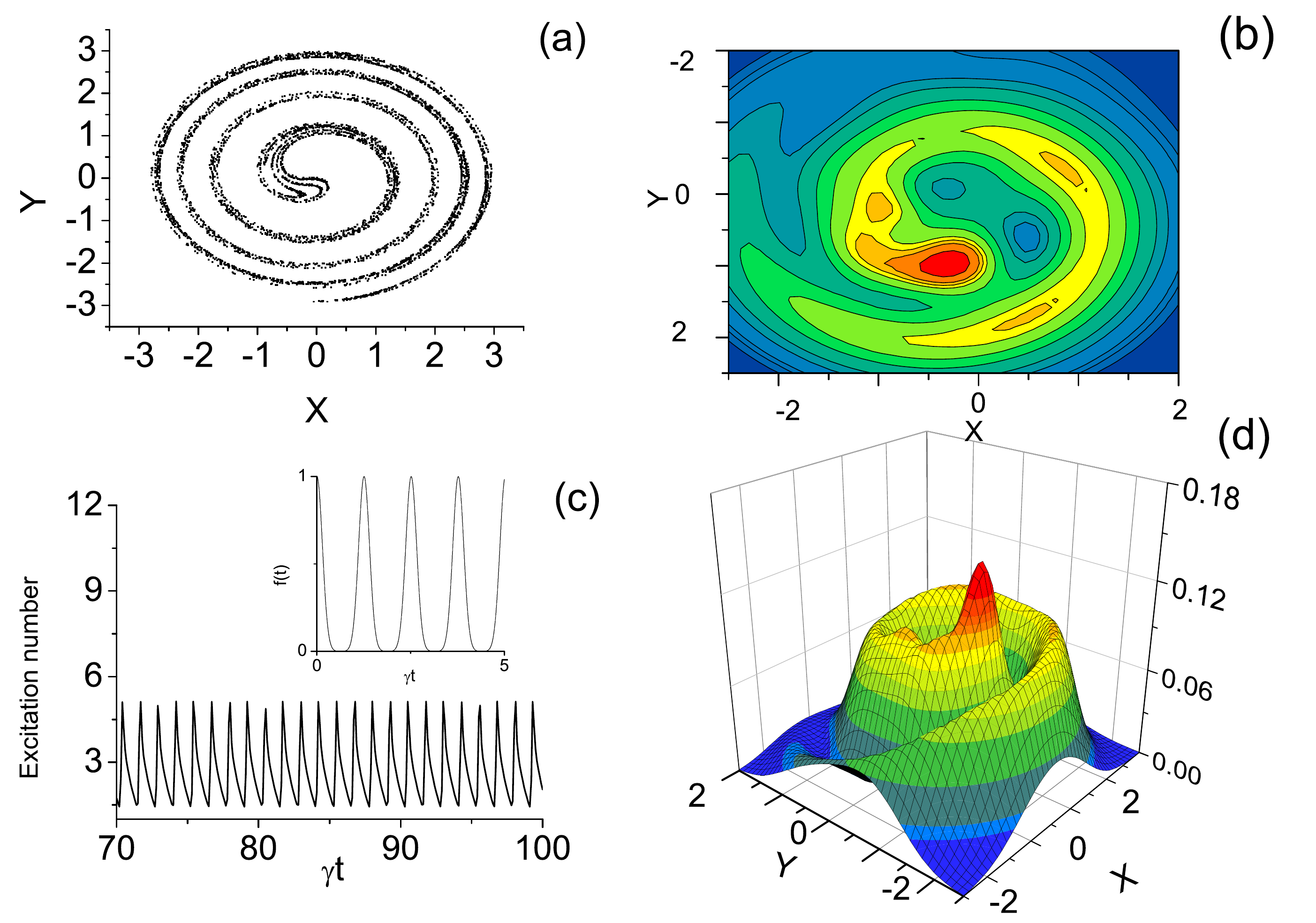}
\caption{(Color online)(a) The Poincar\'e section, (b) the contour plot of Wigner function, (c) the excitation number time series with snapshots of Gaussian pulses as inset, and (d) the Wigner function. The parameters are: $\chi/\gamma=0.7$, $\Omega/\gamma=20.4$, $\Delta/\gamma=-15$, $T=0.15\gamma^{-1}$, $\tau=2\pi/5\gamma$. Note that these distributions occur at time $\gamma t=100$ with the mean excitation number of $2.04$. The range of excitation numbers are $1.43$ to $5.13$.}
\label{pulse_duration3}
\end{figure}

\begin{figure}
\includegraphics[width=8.6cm]{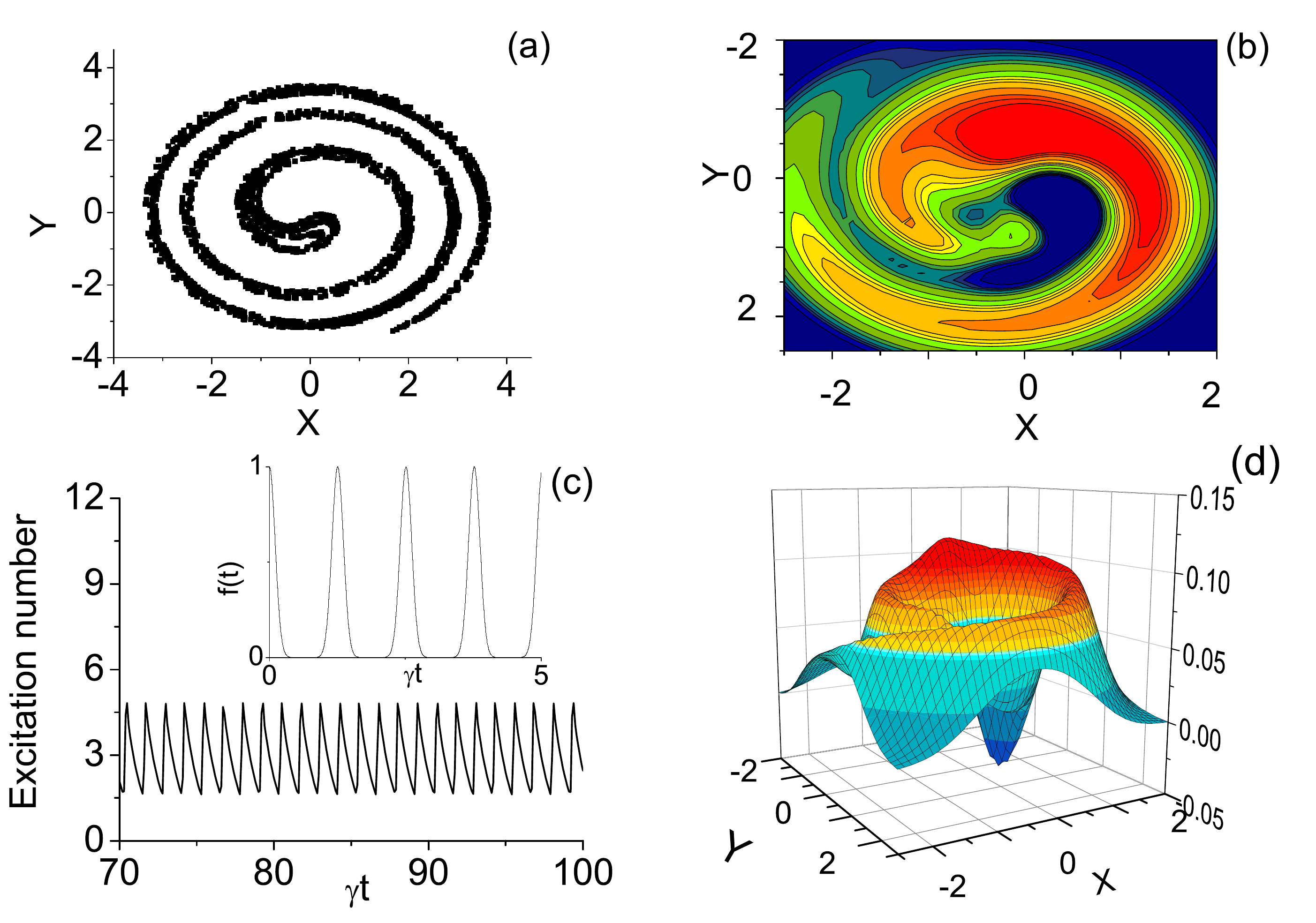}
\caption{(Color online)(a) The Poincar\'e section, (b) the contour plot of Wigner function, (c) the excitation number time series with snapshots of Gaussian pulses as inset, and (d) the Wigner function. The parameters are: $\chi/\gamma=0.7$, $\Omega/\gamma=20.4$, $\Delta/\gamma=-15$, $T=0.1\gamma^{-1}$, $\tau=2\pi/5\gamma$. Note that these distributions occur at time $\gamma t=100$ with the mean excitation number of $2.46$. The range of excitation numbers are $1.62$ to $4.83$.}
\label{pulse_duration4}
\end{figure}

The typical results of calculations are depicted below. Note, that the
ensemble-averaged mean oscillatory excitation number and the Wigner functions
are nonstationary and exhibit a periodic time dependent behavior, i.e. they
repeat the periodicity of the driving pulses at the over transient regime. In
this nonstationary regime the Poincar\'e section depends on the initial
time-interval $t_0$. We choose various initial time $t_0$ in order to ensure
that they match to the corresponding time-intervals of the Wigner function. In
Figs. \ref{pulse_duration1}, \ref{pulse_duration2}, \ref{pulse_duration3} and \ref{pulse_duration4}, we show the typical semiclassical and quantum
distributions for the parameters $\Delta/\gamma $, $\chi/\gamma$,
$\Omega/\gamma$ corresponding to the chaotic regimes and for various durations
of the Gaussian pulses.


As we see, the figures of Poincar\'e sections clearly indicate a classical strange attractor with fractal structure that is typical of chaotic dynamics. The Wigner functions have spiral (helical) structures (Figs. \ref{pulse_duration1} to \ref{pulse_duration4}) that reflect chaotic regime in analogy with the corresponding Poincar\'e sections, and their contour plots are concentrated approximately around the attractor. Nevertheless, in this deep quantum regime the different branches of the attractors are hardly resolved in the Wigner functions.

It should be specified that the Wigner function for the regime presented in
Fig. \ref{pulse_duration4} has region of negative values. Obviously, this fact
reflects on quantum effects in the chaotic regime.

\begin{figure}
\includegraphics[width=8.6cm]{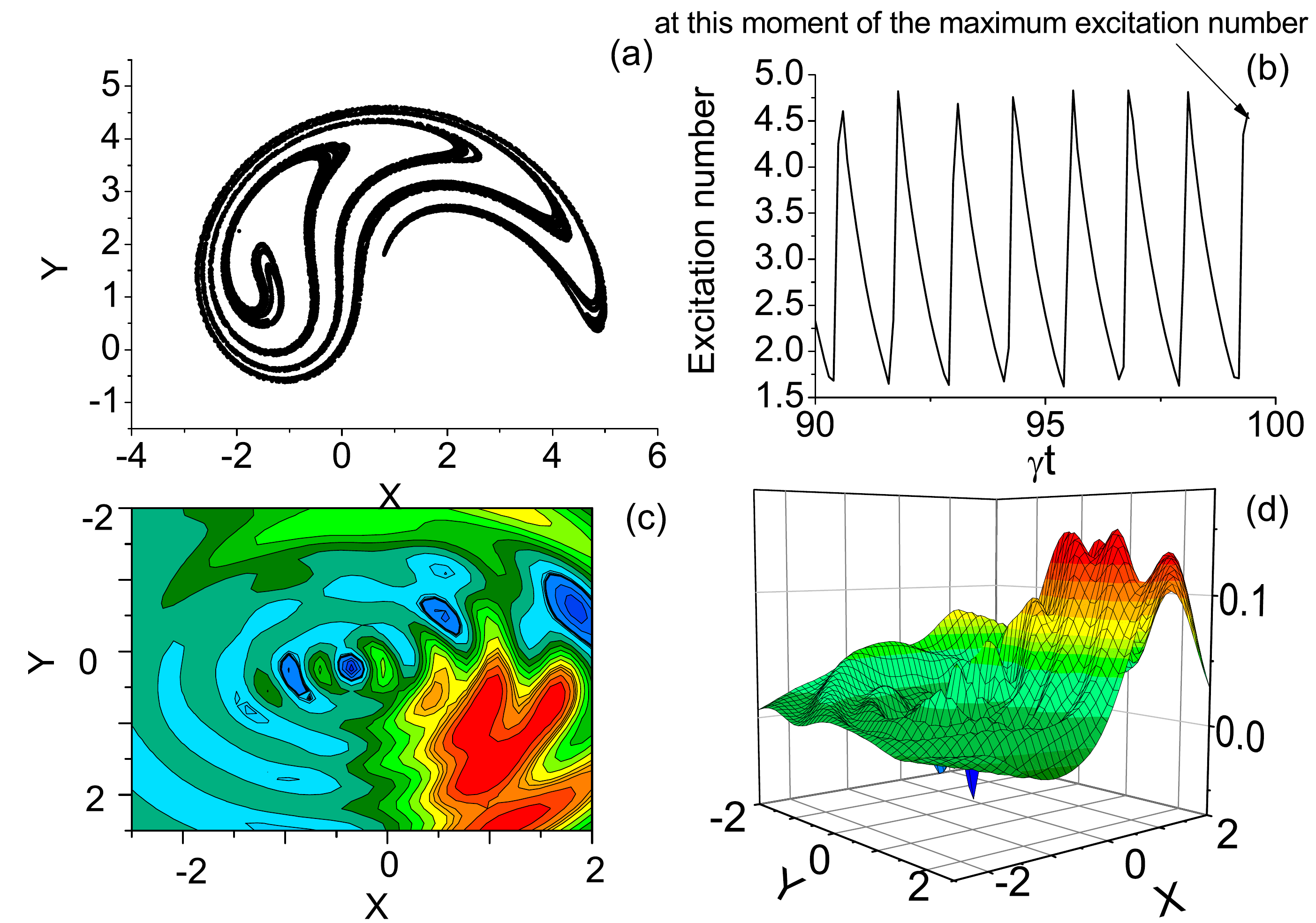}
\caption{(Color online)(a) The Poincar\'e section, (b) the excitation number time series, (c) the contour plot of Wigner function, and (d) the Wigner's function at the moment $\gamma t=100.6$ when the excitation number is maximum. The parameters are: $\chi/\gamma=0.7$, $\Omega/\gamma=20.4$, $\Delta/\gamma=-15$, $T=0.1\gamma^{-1}$, $\tau=2\pi/5\gamma$.}
\label{ex_num_max}
\end{figure}

\begin{figure}
\includegraphics[width=8.6cm]{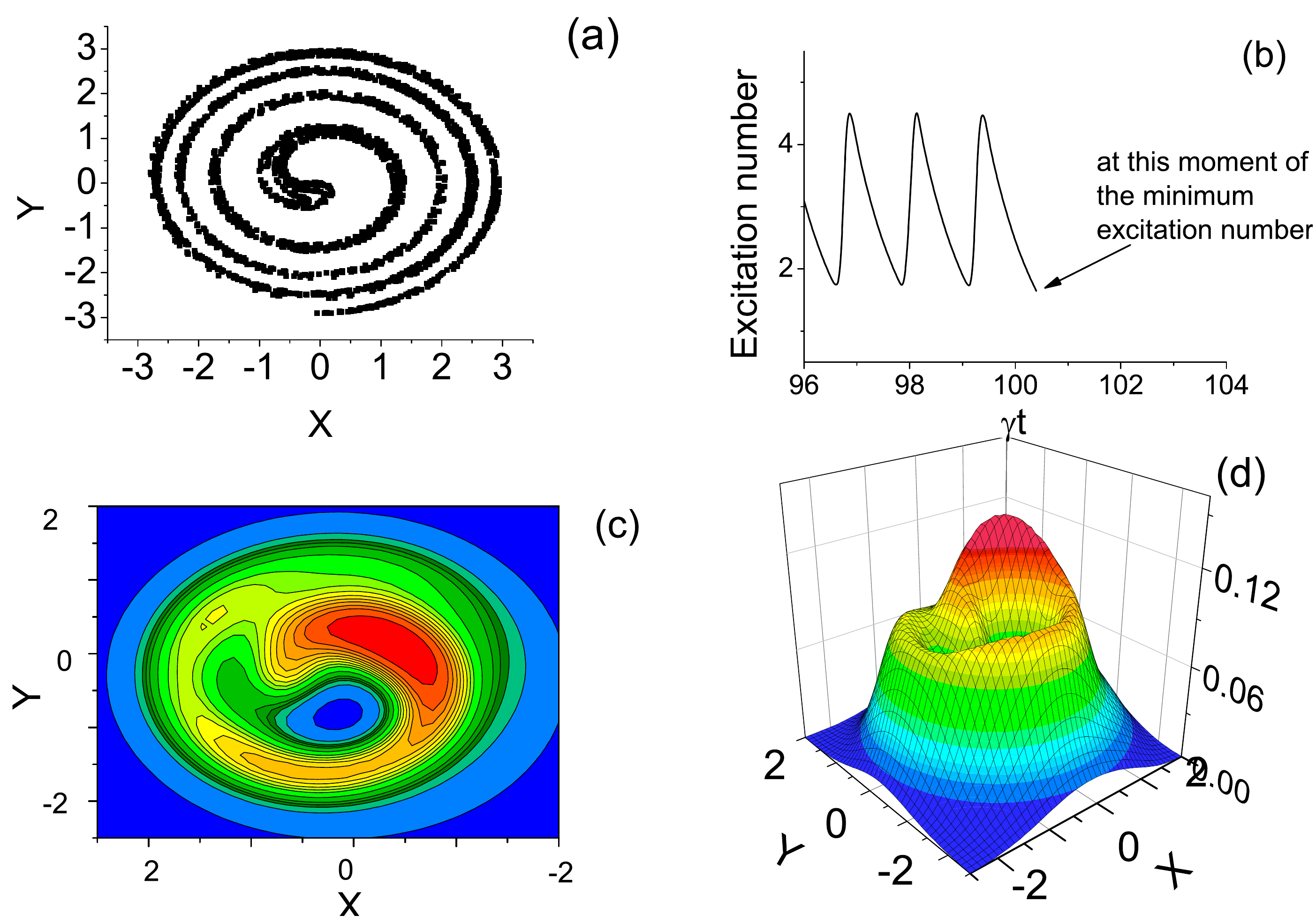}
\caption{(Color online)(a) The Poincar\'e section, (b) the excitation number time series, (c) the contour plot of Wigner function, and (d) the Wigner's function at the moment $\gamma t=100.4$ when the excitation number is minimum. The parameters are as follows: $\chi/\gamma=0.7$, $\Omega/\gamma=20.4$, $\Delta/\gamma=-15$, $T=0.1\gamma^{-1}$, $\tau=2\pi/5\gamma$.}
\label{ex_num_min}
\end{figure}

The chaotic dynamics of the oscillatory mode strongly depends on the
time-interval $t$. To demonstrate this point in Figs. \ref{ex_num_max} and
\ref{ex_num_min}, we depict the Wigner function and the Poincar\'e section for
the oscillatory parameters used in Fig. \ref{pulse_duration4}, however, for the
other time-intervals of $t$ within the duration of pulses which correspond to
the maximal and minimal values of the number of excitation number.

\begin{figure}
\includegraphics[width=8cm]{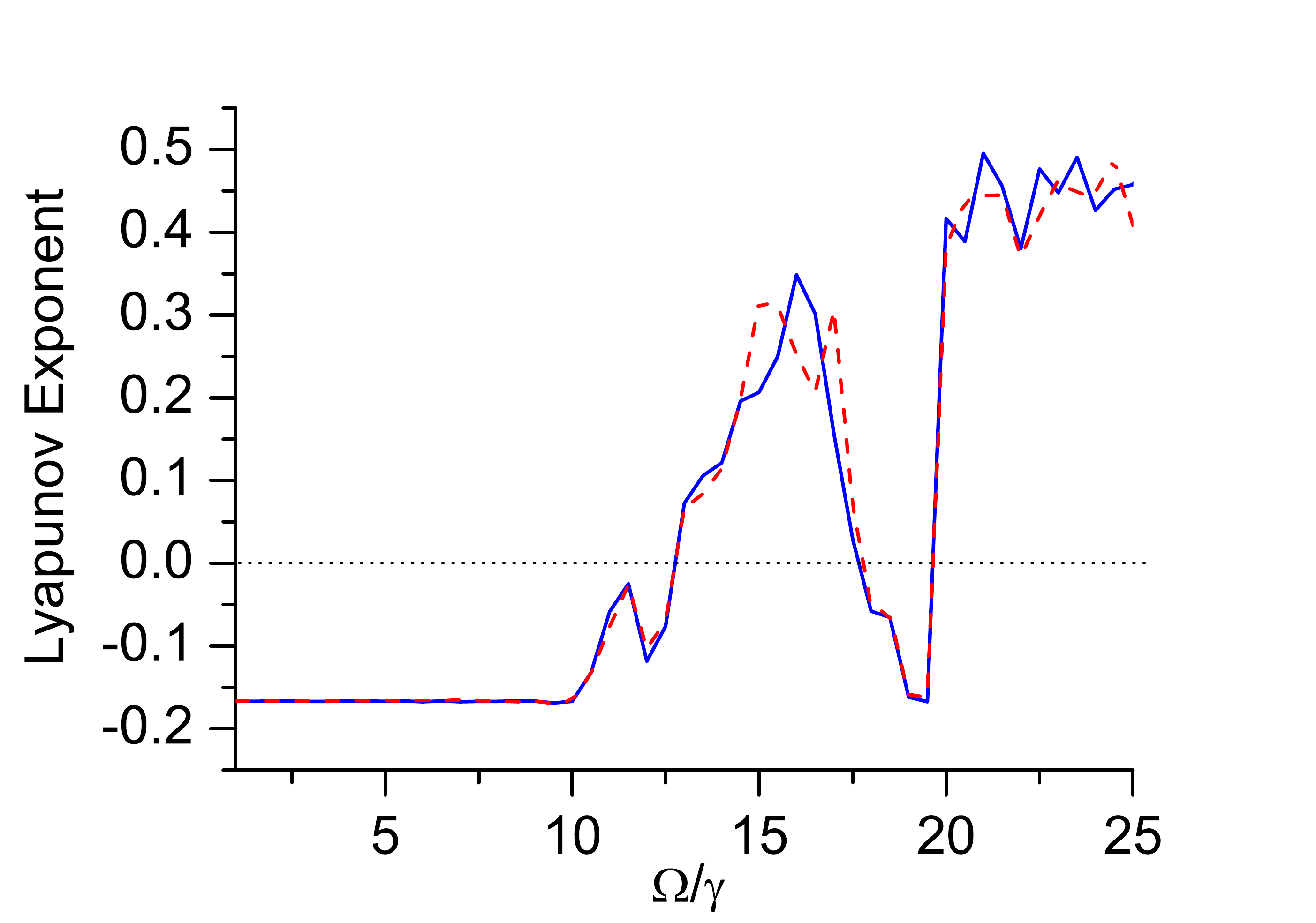}
\caption{(Color online) The largest Lyapunov exponent of the semi-classical dynamics versus
the strength of the pulse trains. The solid (blue) curve corresponds to the
moment of maximum excitation number at $\gamma t=39.1$ for $1 \leq
\Omega/\gamma \leq 19$ and at $\gamma t=39$ for $19.5 \leq \Omega/\gamma \leq
26$. The dashed (red) curve corresponds to the moment of minimum excitation
number at $\gamma t=40.2$ for $1 \leq \Omega/\gamma \leq 8.5$ and at $\gamma
t=40.1$ for $9 \leq \Omega/\gamma \leq 26$. The rest of the parameters are:
$\chi/\gamma=0.7$, $\Delta/\gamma=-15$,
$T=0.1\gamma^{-1}$, $\tau=2\pi/5\gamma$.}
\label{Lya_exp}
\end{figure}

\begin{figure}
\includegraphics[width=8.6cm]{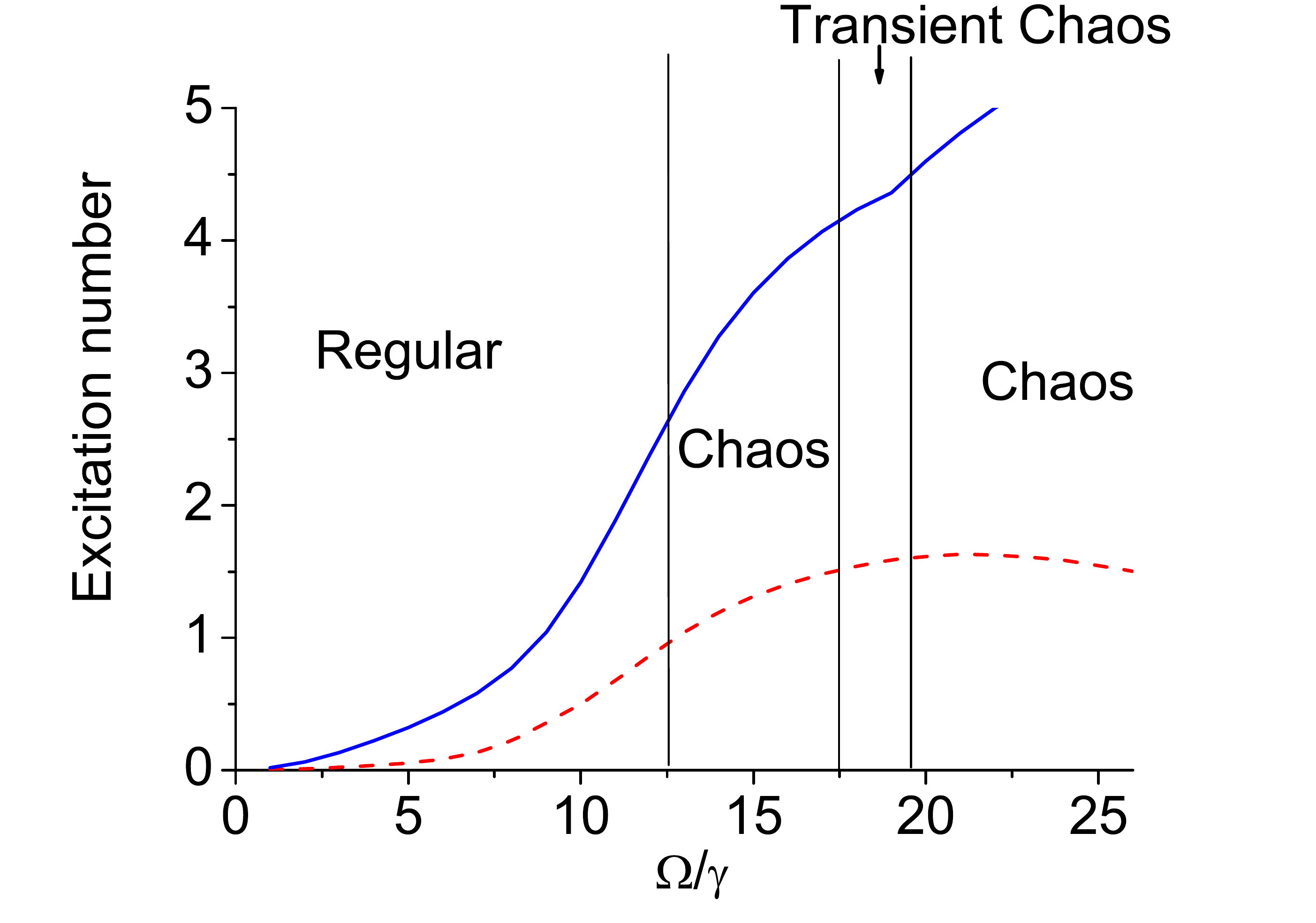}
\caption{(Color online) The plot of the maximum and minimum excitation number versus the
strength of the pulse trains that correspond to the case of Fig. \ref{Lya_exp}.
The solid (blue) curve is for the maximum excitation number and the dashed
(red) curve for the minimum excitation number. Note that the regular, transient
chaos, and chaos regimes of the corresponding semi-classical dynamics are also
indicated in the plot.}
\label{max_min_exnum}
\end{figure}

In the end of this section, we explore the transition from regular to chaotic
regimes that are realized through varying the strength of the pulse trains by
considering cases of small excitation number. For this goal, we present at
first the results in semiclassical approximation based on an analysis of the
Lyapunov exponents of the semi-classical time series \cite{sprott}.
This quantity is determined as
$L=\frac{1}{\Delta t}\ln\frac{\|x_2(t)-x_1(t)\|}{\|x_2(t_0)-x_1(t_0)\|}$,  
here $x=\left(Re(\alpha),Im(\alpha),\beta\right)$, where $\beta$ is the
time variable defined through $d\beta/dt=1$ which augments 
Eq. (\ref{semclass}) to create an autonomous system.
Note that $x_2$ and $x_1$ represent two trajectories that are very close 
together at the initial time $t_0$. Furthermore, $\Delta t = t - t_0$, with 
$t \rightarrow \infty$. For $L>0$ the system shows chaotic 
dynamics. $L=0$ corresponds to the case of 
conservative regular systems, and $L<0$ indicates that the dissipative system 
is regular. We examine the exponents for time intervals corresponding to 
the minimal and maximal excitation numbers of the oscillatory mode in 
dependence from the parameter $\Omega/\gamma$. The results are depicted in 
Figs. \ref{Lya_exp} and \ref{max_min_exnum} for the constant parameters:
$\Delta/\gamma $ and $\chi/\gamma$. We observe a transition from regular to
chaotic behavior at $\Omega/\gamma=12.55$ for minimal and maximal excitation
number of the oscillatory mode. Note that this transition occurs at the strong
quantum regime as the excitation number ranges from the minimum $n=0.94$ to the
maximum $n=2.70$. Thus, when the strength of the pulse trains $\Omega/\gamma$
is low, we observe regular behaviour; and as a critical threshold is crossed,
the system behaves chaotically. Interestingly, a closer scrutiny of the
dynamics of the system reveals a regime of transient chaos in the range $17.61
\leq \Omega/\gamma \leq 19.56$, whereupon the semi-classical dynamics rattle
about chaotically for some time before settling down to regular behaviour which
leads to a window of negative Lyapunov exponents. Then, beyond
$\Omega/\gamma=19.56$, chaotic attractors are found to emerge again in the
Poincar\'e section. It is important to note that analogous dynamical behavior
are observed to arise at both the moments when the excitation number is a
minimum and a maximum (see Fig. \ref{Lya_exp}).

In summary, we have uncovered the parameters for which decreasing of the
excitation number leads to a transition of the system from chaotic to regular
regime in the classical treatment. Another situation is realized in the quantum
treatment. Indeed, for these regimes of low-excitation, quantum noise and
quantum effects play essential role in forming of the oscillatory dynamics,
particularly, in realization of chaotic dynamics and scenarios of transition
from chaotic to regular regime. This statement is also confirmed by the
calculation of the Wigner function. We observe that below the transition
threshold of $\Omega/\gamma=12.55$, the Wigner function has the form of a
single hump. Beyond the transition threshold, the hump starts to spread and
spiral begins to form. These changes are found to happen continuously and
smoothly in the Wigner function as $\Omega/\gamma$ increases. However, as
$\Omega/\gamma$ goes above $19.56$, the Wigner functions are observed to
quickly take the appearance of a strange attractor. Our results here thus
clearly show the presence of good quantum-classical correspondence in the
quantum and classical dynamical behaviour. In addition, our results also
demonstrate that a quantum NDO, which is a form of quantum anharmonic
oscillators, possess a rich set of dynamical behaviour and properties that may
be potentially useful for many practical purposes \cite{cnn_pre, cnn_pra}.

\section{Summary}
To summarize, we have studied the problem of quantum chaos and bistability at
level of few excitation numbers of NDO that is interacting with an external
field  and a background environment leading to dissipation and decoherence. We
have analyzed the formation of bistable behavior and chaotic regime of NDO in
strong quantum regime when the ratio $\chi/\gamma$ is chose from $0.7$ to  $2$
. We use a systematic numerical analysis  based on numerical simulation of
master equation by using quantum state diffusion method  of quantum
trajectories where we have varied a number of relevant oscillatory parameters:
$\Omega/\gamma$, $\Delta/\gamma$ as well as the parameters of Gaussian
pulses. Thus, the combination of
the results provides a rather thorough understanding of the NDO in strong
quantum regime as well as have presented a simple picture to understand the
formation of bistability and chaos at low-level of quanta in phase-space. We
have also found unexpected features of NDO in phase-space. It has been
demonstrated that the Wigner functions of oscillatory mode in both bistable and
chaotic regimes realized due to interaction with train of Gaussian pulses
acquire negative values and interference patterns in parts of phase-space.  We
have demonstrated that in the case of bistable dynamics the Wigner functions
describe two humps corresponding to bistable branches and interference pattern
between them in phase-space while for the chaotic regime the Wigner functions
have spiral  structures, (which  correspond to strange  attractor in
Poincar\'e sections), with deep well showing negativity of the Wigner function
(Fig. \ref{pulse_duration4} (d)). Quantum interference in phase-space is
realized  in over transient regime due to driving the oscillator by series of
short pulses with proper parameters for effective reducing of dissipative and
decoherence effects. Our results can be tested with available experimental
systems noted in the section \ref{intro}.

\end{document}